\documentclass[fleqn,usenatbib]{mnras}

\usepackage{newtxtext,newtxmath}

\usepackage[T1]{fontenc}

\DeclareRobustCommand{\VAN}[3]{#2}
\let\VANthebibliography\thebibliography
\def\thebibliography{\DeclareRobustCommand{\VAN}[3]{##3}\VANthebibliography}

\usepackage{graphicx}	
\usepackage{amsmath}	
\usepackage{tabularx}
\usepackage{supertabular}

\title{Search for Interpulses in a Complete Sample of Pulsars
at a Frequency of 111 MHz}
\author[M. O. Toropov et al.]{
M. O. Toropov $^{1}$,\thanks{E-mail: }
S. A. Tyul'bashev,$^{2}$
V.S. Beskin $^{3,4}$
\\
$^{1}$ LLC TEK Inform, Moscow, 117246 Russia \\
$^{2}$ Pushchino Radio Astronomy Observatory, Astro Space Center, Lebedev Physical Institute, Russian Academy of Sciences, Pushchino, 142290 Russia \\
$^{3}$ Lebedev Physical Institute, Moscow, Russia \\
$^{4}$ Moscow Institute of Physics and Technology (MIPT), Dolgoprudnyi, Russia \\
}

\date{2024}

\pubyear{2024}

\begin{document}
\label{firstpage}
\pagerange{\pageref{firstpage}--\pageref{lastpage}}
\maketitle

\begin{abstract}
An interpulse search was carried out in a sample of 96 pulsars observed on the Large Phased Array (LPA) radio telescope in the Pushchino Multibeams Pulsar Search (PUMPS). The pulsar sample is complete for pulsars having a signal-to-noise ratio (S/N) in the main pulse ($MP$) greater than 40. To search for weak interpulses ($IP$), the addition of average profiles over an interval of up to 10 years was used. Interpulses were detected in 12 pulsars (12.5\% of the sample), of which 7 pulsars have an interpulse located near the $180^{\circ}$ phase relative to the main pulse (probable orthogonal rotators), and 5 have phases far from $180^{\circ}$ (probable coaxial rotators). The amplitude ratios of $IP/MP$ are in the range of 0.004-0.023, the median value is 0.01. Estimates of the observed number of coaxial and orthogonal rotators with $IP/MP\ge 0.01$ do not contradict the model according to which, during evolution, the magnetic axis and the axis of rotation become orthogonal.
\end{abstract}

\begin{keywords}
pulsar, interpulse, coaxial (alignment) rotator, orthogonal (counter-alignment) rotator
\end{keywords}



\section{Introduction}

As is known, radio pulsars are rapidly rotating neutron stars, the pulsed radiation of which we see due to the non-zero angle between the axis of rotation and the axis of the magnetic dipole. If, during the rotation of a neutron star, a cone of radiation enters the beam of view, along the central axis of which there is a line connecting the magnetic poles, we can observe a flash of radio emission. According to the ATNF catalog (https://www.atnf.csiro.au/research/pulsar/psrcat/;  \citeauthor{Manchester2005}, \citeyear{Manchester2005}), the typical half-width of the pulse in fractions of the period is 1-10\%, which allows us to roughly estimate the width of the cone as $3^{\circ}-30^{\circ}$.

It is obvious that the observed half-width of the pulse depends not only on the angle between the axis of rotation and the axis of the magnetic field, but also on the angle at which the central line in the cone of radiation is located in relation to the line of sight. For example, if the angle between the axes is small (coaxial rotator), the radiation cone can capture the axis of rotation and the observed pulse can occupy the entire period. There is a possibility that such a radio pulsar will go unnoticed in ordinary pulsar observations, when we add up many pulses with a known period and then look for a peak in the average profile. Another extreme case is when the axis of rotation and the axis of the magnetic field are orthogonal (orthogonal rotator). In this case, if the axis of rotation is orthogonal to the line of sight, we will observe both magnetic poles.

Pulsars in which two pulses are observed at each revolution (period) of the pulsar are called pulsars with interpulses. They can be both coaxial and orthogonal rotators (\citeauthor{Gil1983}, \citeyear{Gil1983}). For coaxial rotators, an interpulse may appear if the detected radiation is generated in two local zones located at different distances from the polar cap.

For an ideal orthogonal rotator, the observed pulses must have close observed flux densities. However, in practice, for known orthogonal rotators, the observed flux densities of the main pulse and interpulse may differ several times (see Table 1 in \cite{Maciesiak2011}). The stronger pulse is called the main pulse ($MP$), the weaker pulse is called the interpulse ($IP$). Determining an orthogonal rotator in practice is a difficult task. In this paper, we will adhere to the empirical approach, according to which the phase distance between $MP$ and $IP$ close to $180^o$ is inherent in orthogonal rotators, and far from 180$^o$ is inherent in coaxial rotators. According to Table 1 in \citeauthor{Maciesiak2011} (\citeyear{Maciesiak2011}), it can be seen that the median value of the phase distance for orthogonal rotators is $180^o$. For coaxial rotators, the median value of the phase distance between $MP$ and $IP$ differs from $180^o$ by $35^o$. At the same time, it is impossible to exclude cases when in a coaxial rotator the phase distance between $MP$ and $IP$ will be close to $180^o$, and in an orthogonal rotator the phase distance between $MP$ and $IP$ will be far from $180^o$.

Geometry determines the necessary conditions for the pulse to be visible, however, there are also physical conditions that determine whether a radio pulse will occur. According to modern concepts \citeauthor{Lorimer2012} (\citeyear{Lorimer2012}), a neutron star will be visible as a radio pulsar if the conditions necessary for the birth of secondary electron-positron plasma are met in the polar regions. The set of physical parameters that determine the existence of radio emission are the period of the pulsar $P$, as well as the magnitude and geometry of the magnetic field.

As the neutron star loses rotational energy to generate pulsar wind, its rotational velocity gradually decreases, and the period $P$ increases accordingly. When the rotation slows down, conditions occur when the neutron star ceases to be a radio pulsar. In the diagram $P/\dot P$ such radio pulsars are approaching the line of death.

In addition to slowing down the rotation of a neutron star, there is also an evolution of the angle between the axis of rotation and the magnetic axis. That is, there are no pulsars in which this angle remains unchanged after their birth. At the same time, according to the works of \citeauthor{Davis1970} (\citeyear{Davis1970}); \citeauthor{Goldreich1970} (\citeyear{Goldreich1970}); \citeauthor{Manchester1977} (\citeyear{Manchester1977}); \citeauthor{Philippov2014} (\citeyear{Philippov2014}), during evolution the axes become co-directional, and according to the works of \citeauthor{Gurevich1993} (\citeyear{Gurevich1993}); \citeauthor{Lyne2015} (\citeyear{Lyne2015}) the axes become orthogonal. That is, purely statistically, we should observe either an excess of coaxial rotators or an excess of orthogonal rotators.

In this paper, a statistical check of the number of observed pulsars with interpulses in a sample of pulsars observed in the Pushchino Multibeam Pulsar Search (PUMPS) was carried out.

\section{Observations and primary data processing}
\begin{figure*}
	\begin{center}
		\includegraphics[width=0.8\textwidth]{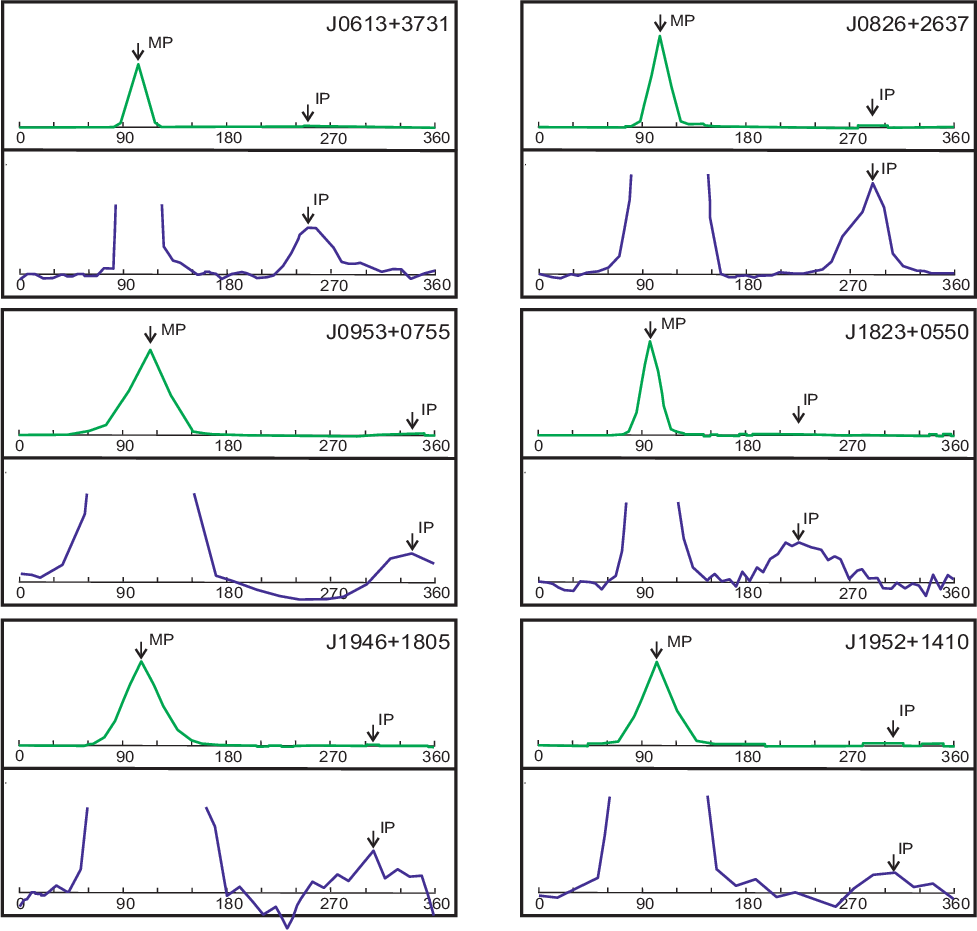}
		\caption{The figure shows 12 paired images, each of which shows the summed average profile of the pulsar in the upper part, and the top of the $MP$ in the profile is removed in the lower part so that the $IP$ is visible. The arrows in the figures indicate the position of $MP$ and $IP$.}
		\label{fig:fig1A}
	\end{center}
\end{figure*}
In August 2014, monitoring observations began on the multibeams radio telescope Large Phased Array (LPA3) of the Lebedev Physical Institute (LPI). This radio telescope appeared during the major reconstruction of the LPI radio telescope, when the technical possibility of creating four independent radio telescopes was made on the basis of one antenna field. The LPA1 radio telescope is used for conventional single-beam observations of pulsars at declinations from $-16^{\circ}$ to $+87^{\circ}$. LPA3 does not have the ability to switch beams, and initially it was assumed that the radio telescope would work according to the program "Space Weather" (\citeauthor{Shishov2016}, \citeyear{Shishov2016}). Its 128 stationary beams are lined up in the meridian plane and cover declinations from $-9^{\circ}$ to $+55^{\circ}$ with beam overlap at the 0.4 level. The beam size is approximately $0.5^{\circ} \times 1^{\circ}$, the time of passage through the half-power meridian (observation session) is about 3.5 minutes. Since the LPA is a meridian instrument, any source in the sky is registered only once a day. One of the tasks of the LPA3 radio telescope is to search for pulsars and transients (\citeauthor{Tyulbashev2016}, \citeyear{Tyulbashev2016}). Observations on LPA3 are carried out in the 2.5 MHz band, divided into 32 frequency channels with a width of 78 kHz. The sampling of the point is fixed and is 12.5 ms.

LPA3 was built for the "Space Weather" project (https://sw.prao.ru/) and it did not include the ability to control the recording time of each data point according to the atomic frequency standard. Observations are carried out in hourly portions. The beginning of each hour (the start of observations) is controlled by a special GPS receiver, and the time inside the hour is counted by a quartz oscillator. Possible time errors by the end of the hour interval may be $\pm 25$~ms. However, until now, all attempts to make a timing that takes into account the assumed accuracy of the quartz oscillator have been unsuccessful.

Every four hours, the LPA dipole lines are disconnected for 15 seconds, and a calibration signal (calibration step) of a known temperature from a noise generator is supplied to the cable paths running from the end of the dipole lines to the entrance to the recorders. The form of the applied signal is OFF-ON-OFF, where OFF corresponds to the disconnection of the dipole lines for 5 seconds. At this time, the recorder records a signal corresponding to the ambient temperature. The ON signal appears when the noise generator amplifiers are connected at the input. It also lasts 5 seconds. The calibration steps reflect the current gain in the paths of the radio telescope, and allows you to align the data received at different times in different frequency channels (\citeauthor{Tyulbashev2019}, \citeyear{Tyulbashev2019}).

The survey has been conducted around the clock for more than 9 years, and more than 3,000 observation sessions have been accumulated for each pulsar entering the covered area. To check the number of observed pulsars with $IP$, we will add up the average profiles for the available days, improving the signal-to-noise ratio ($S/N$) in $MP$, which we will place at a quarter distance from the beginning of the average profile, and check the appearance of statistically significant signals (probable $IP$) on different distances from $MP$.

As mentioned above, timing based on monitoring data is not yet possible, and therefore, to increase $S/N$, you can add up the average profiles of only those pulsars that are visible in one observation session. The search for pulsars on LPA3 shows that about 300 second pulsars are recorded in monitoring observations (https://bsa-analytics.prao.ru/pulsars/), however, not all of these pulsars are regularly observed with $S/N$ sufficient to see their average profiles, and for some of these pulsars, profiles have not been obtained at all (\citeauthor{Tyulbashev2022}, \citeyear{Tyulbashev2022}).

To search for interpulses, we selected only those pulsars whose main pulse was greater than 6 in the observation session $S/N$. Visually, it is easy to see pulses with $S/N = 4-5$ on the recordings, but an attempt to add average profiles with such $S/N$ in automatic mode led to the appearance of artifacts in the summed average profiles. As shown in \citet{Tyulbashev2024}, the maximum sensitivity when searching for pulsars in monitoring data is achieved for pulsars having P>0.25s, $DM < 100$~pc/cm$^3$ located in the site $+21^{\circ} < \delta < +42^{\circ}$. We searched for interpulses for pulsars previously detected in PUMPS and having P>0.25s.

According to the works of \citeauthor{Maciesiak2011} (\citeyear{Maciesiak2011}); \citeauthor{Malov2013} (\citeyear{Malov2013}) of the 44 pulsars with known $IP$, the ratio of $IP/MP\ge $ is 0.1 for almost 90\%, namely for 39 pulsars out of 44. This means that if you set the criterion $S/N > 40$ for $MP$, $IP$ will turn out to be $S/N > 4$ for 90\% of pulsars, if the pulsar has $IP$. In this case, the signals $S/N > 4$ are visually detected in the records of the average profiles. Therefore, in addition to the condition $S/N > 6$ for momentum in a separate session, we also set the condition $S/N > 40$ in the accumulated average profile. Naturally, for strong pulsars, much weaker $IP$ can also be detected. The condition $S/N > 40$ indicates what minimum fraction of pulsars having $IP$ can be guaranteed to be detected. This condition also allows you to fix the completeness of the sample for the observed $S/N$.

When processing observations, the following procedures were performed:

- normalization of raw data by calibration step;

- addition of frequency channels taking into account a known dispersion measure ($DM$) and subsequent addition of the resulting series taking into account a known period ($P$) of the pulsar to obtain an average profile;

- if $MP$ has its $S/N > 6$, profiles are stored in temporary tables (metadata is also stored: date and hour of observations in UTC; number of the point in the hourly file from which data cutting begins; number of the maximum point in the profile after adding all pulses; pulse height in units of $S/N$);

- the standard deviations of noise ($\sigma_n$) outside the pulse are estimated and normalized to the obtained value (with this approach, $\sigma_n = 1$ in each selected session, and therefore it is possible to add average profiles for different days without loss of $S/N$);

- an annular shift is performed in each selected average profile until the main pulse reaches a value of about $90^{\circ}$ on the phase representation of the profile;

- we add up the selected average profiles for each pulsar for all years, and leave for analysis only those pulsars with $S/N > 40$ in the summed average profile;

- we are looking for statistically significant values ($S/N > 4 \sigma_n$) of new peaks (interpulses) on the average profile (since the width of $IP$ is comparable to the width of $MP$, we are not looking for single-point outliers, but structures comparable in width to $MP$);

- we conduct a visual check of the candidates found. 

As a result of the processing done, we get tables that contain average pulsar profiles summarized separately by year. We also have profiles summarized over the years. All this allows you to view any of the studied pulsars in any detail, including individual observation sessions.

\section{Results and analysis}

After processing the observations, 96 pulsars remained, having $S/N > 40$ in $MP$ after summing the average profiles. In table 1, we place information on these pulsars. Columns 1-3 show the name of the pulsar in the J2000 agreement (according to the ATNF catalog), its $P$ and $DM$, columns 4 and 5 show the ratio $IP/MP$ and the distance between $MP$ and $IP$ in degrees. For all $IP$ found, the maximum value of $S/N > 4$, the duration of $IP$ is comparable to the duration of $MP$. The accuracy of determining the phase of $IP$ is low, especially for pulsars with short periods. Since the time between counts is 12.5 ms, the pulsar with a period of 0.25s has an entire average profile of 20 points. At the same time, due to the peculiarities of pulse addition, when the apparent maximum of $MP$ is considered to be the exact location of the pulse center, the location error may be $\pm 1$ point from the true value. The position of the maximum on $IP$ can also be determined with the error $\pm 1$ point or more (see Fig.~\ref{fig:fig1A}). Finally, we give an error to determine the phase of $IP$ as $\pm 1.5$ points and convert this error to degrees, taking into account the known period. Note also that in this paper, the position of $IP$ is given relative to $MP$, and the distance is determined as in Fig.1, when degrees are counted from zero, $MP$ is in the phase of $90^o$, and the last point of the average profile corresponds to the phase of $360^{\circ}$. For example, for the pulsar J0953+0755, we have marked the position of $IP$ at $216^{\circ} \pm 27^{\circ}$. The work \citeauthor{Gil1983} (\citeyear{Gil1983}) says that $IP$ precedes $MP$ and the distance to it is $152^{\circ}$. Since $360^{\circ} - 216^{\circ} = 144^{\circ}$ - this means that within the specified errors, the positions of $IP$ on the average profiles coincide.
\begin{figure*}
	\begin{center}
		\includegraphics[width=0.8\textwidth]{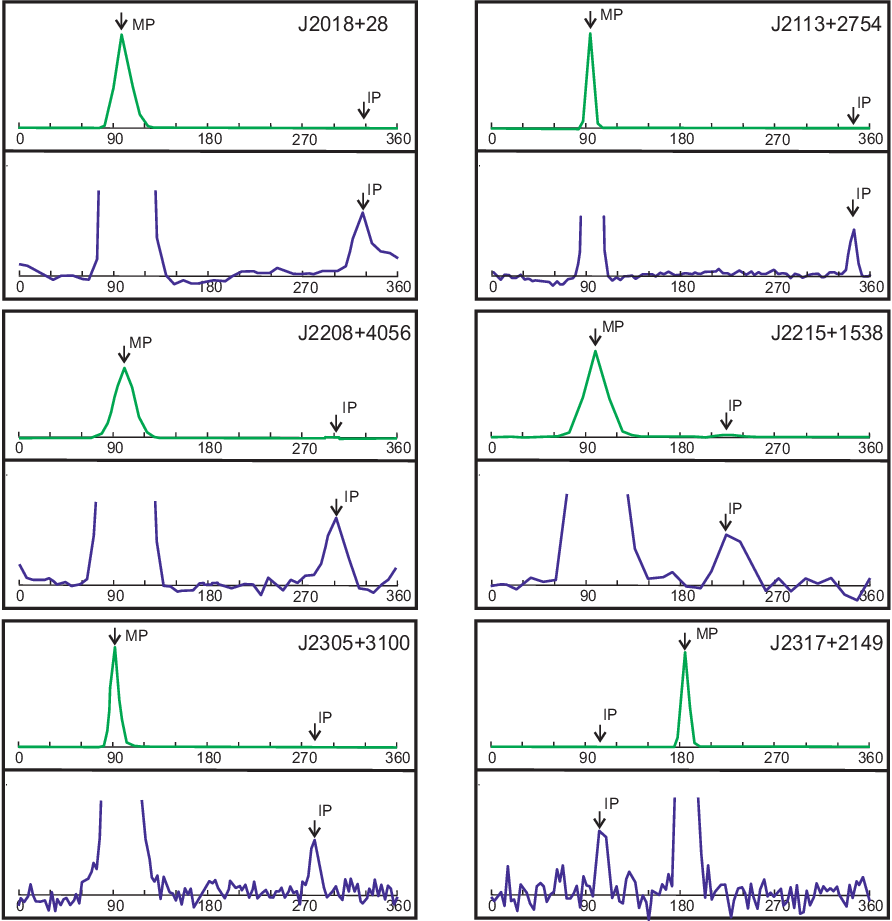}
		\caption{The figure shows 12 paired images, each of which shows the summed average profile of the pulsar in the upper part, and the top of the $MP$ in the profile is removed in the lower part so that the $IP$ is visible. The arrows in the figures indicate the position of $MP$ and $IP$.}
		\label{fig:fig1B}
	\end{center}
\end{figure*}

\begin{table*}
	\centering
	\caption{The ratio of $IP/MP$ for 96 pulsars studied}
	\label{tab:tab1}
	\begin{tabular}{c|c|c|c|c||c|c|c|c|c|}
		\hline
		Name & $P$(s) & $DM$ (pc/cm$^3$) & $IP/MP$ & r($^o$) & Name & $P$(s) & $DM$ (pc/cm$^3$) & $IP/MP$ & r($^o$)\\
		\hline
		J0014+4746	&	1.2406	&	30.4	&	<0.01	&	-	&	J1635+2418	&	0.4905	&	24.2	&	<0.002	&	-	\\
		J0034-0721	&	0.9429	&	10.9	&	<0.0005	&	-	&	J1638+4005	&	0.7677	&	33.4	&	<0.015	&	-	\\
		J0039+35	&	0.5367	&	53	&	<0.005	&	-	&	J1645+1012	&	0.4108	&	36.1	&	<0.003	&	-	\\
		J0048+3412	&	1.217	&	39.9	&	<0.003	&	-	&	J1645-0317	&	0.3876	&	35.7	&	<0.006	&	-	\\
		J0051+0423	&	0.3547	&	13.9	&	<0.01	&	-	&	J1649+2533	&	1.0152	&	34.4	&	<0.009	&	-	\\
		J0137+1654	&	0.4147	&	26	&	<0.019	&	-	&	J1740+1311	&	0.803	&	48.6	&	<0.001	&	-	\\
		J0304+1932	&	1.3875	&	15.6	&	<0.0008	&	-	&	J1741+2758	&	1.3607	&	29.1	&	<0.003	&	-	\\
		J0317+13	&	1.9743	&	12.9	&	<0.013	&	-	&	J1752+2359	&	0.409	&	36.1	&	<0.010	&	-	\\
		J0323+3944	&	3.032	&	26.1	&	<0.004	&	-	&	J1758+3030	&	0.9472	&	35	&	<0.003	&	-	\\
		J0332+5434	&	0.7145	&	26.7	&	<0.002	&	-	&	J1813+4013	&	0.931	&	41.5	&	<0.003	&	-	\\
		J0335+4555	&	0.2692	&	47.1	&	<0.017	&	-	&	J1821+4147	&	1.2618	&	40.6	&	<0.012	&	-	\\
		J0454+4529	&	1.3891	&	20.8	&	<0.005	&	-	&	J1823+0550	&	0.7529	&	66.7	&	0.008	&	$140 \pm 9$	\\
		J0459-0210	&	1.133	&	21	&	<0.005	&	-	&	J1838+1650	&	1.9019	&	32.9	&	<0.010	&	-	\\
		J0525+1115	&	0.3544	&	79.4	&	<0.019	&	-	&	J1841+0912	&	0.3813	&	49.1	&	<0.004	&	-	\\
		J0528+2200	&	3.7455	&	50.8	&	<0.0009	&	-	&	J1844+1454	&	0.3754	&	41.4	&	<0.0008	&	-	\\
		J0540+3207	&	0.5242	&	61.9	&	<0.011	&	-	&	J1849+2559	&	0.5192	&	75	&	<0.063	&	-	\\
		J0546+2441	&	2.8438	&	73.8	&	<0.010	&	-	&	J1851+1259	&	1.2053	&	70.6	&	<0.009	&	-	\\
		J0611+30	&	1.412	&	45.2	&	<0.002	&	-	&	J1907+4002	&	1.2357	&	30.9	&	<0.003	&	-	\\
		J0612+3721	&	0.2979	&	27.1	&	<0.005	&	-	&	J1912+2525	&	0.6219	&	37.8	&	<0.004	&	-	\\
		J0612+37215	&	0.4438	&	39.2	&	<0.012	&	-	&	J1920+2650	&	0.7855	&	27.7	&	<0.004	&	-	\\
		J0613+3731	&	0.6191	&	18.9	&	0.019	&	$155 \pm 11$	&	J1921+2153	&	1.3373	&	12.4	&	<0.0001	&	-	\\
		J0623+0340	&	0.6137	&	54	&	<0.019	&	-	&	J1922+2110	&	1.0779	&	217	&	<0.011	&	-	\\
		J0629+2415	&	0.4766	&	84.1	&	<0.002	&	-	&	J1929+1844	&	1.2204	&	112	&	<0.023	&	-	\\
		J0659+1414	&	0.3849	&	13.9	&	<0.018	&	-	&	J1946+1805	&	0.4406	&	16.1	&	0.011	&	$195 \pm 16$	\\
		J0811+37	&	1.2483	&	16.9	&	<0.008	&	-	&	J1952+1410	&	0.275	&	31.5	&	0.006	&	$188 \pm 25$	\\
		J0815+4611	&	0.4342	&	11.2	&	<0.015	&	-	&	J1954+2923	&	0.4266	&	7.9	&	<0.005	&		\\
		J0826+2637	&	0.5306	&	19.4	&	0.015	&	$180 \pm 13$	&	J2007+0910	&	0.4587	&	48.7	&	<0.018	&		\\
		J0837+0610	&	1.2737	&	12.8	&	<0.0003	&	-	&	J2017+2043	&	0.5371	&	60.4	&	<0.087	&		\\
		J0922+0638	&	0.4306	&	27.2	&	<0.0005	&	-	&	J2018+2839	&	0.5579	&	14.1	&	0.004	&	$224 \pm 12$	\\
		J0935+33	&	0.9615	&	18.3	&	<0.008	&	-	&	J2022+2854	&	0.3434	&	24.6	&	<0.0008	&	-	\\
		J0943+2253	&	0.5329	&	27.2	&	<0.006	&	-	&	J2037+1942	&	2.0743	&	36.8	&	<0.010	&	-	\\
		J0944+4106	&	2.2294	&	21.4	&	<0.007	&	-	&	J2046+1540	&	1.1382	&	39.8	&	<0.006	&	-	\\
		J0946+0951	&	1.0977	&	15.3	&	<0.003	&	-	&	J2055+2209	&	0.8151	&	36.3	&	<0.005	&	-	\\
		J0953+0755	&	0.253	&	2.96	&	0.022	&	$216 \pm 27$	&	J2113+2754	&	1.2028	&	25.1	&	0.007	&	$247 \pm 6$	\\
		J1115+5030	&	1.6564	&	9.18	&	<0.004	&	-	&	J2116+1414	&	0.4401	&	56.2	&	<0.006	&	-	\\
		J1136+1551	&	1.1879	&	4.84	&	<0.001	&	-	&	J2124+1407	&	0.694	&	30.2	&	<0.017	&	-	\\
		J1238+2152	&	1.1185	&	17.9	&	<0.005	&	-	&	J2139+2242	&	1.0835	&	44.1	&	<0.009	&	-	\\
		J1239+2453	&	1.3824	&	9.25	&	<0.001	&	-	&	J2157+4017	&	1.5252	&	71.1	&	<0.002	&	-	\\
		J1246+2253	&	0.4738	&	17.7	&	<0.013	&	-	&	J2208+4056	&	0.6369	&	11.8	&	0.014	&	$198 \pm 11$	\\
		J1313+0931	&	0.8489	&	12	&	<0.008	&	-	&	J2209+22	&	1.7769	&	46.3	&	<0.024	&	-	\\
		J1404+1159	&	2.6504	&	18.4	&	<0.007	&	-	&	J2212+2933	&	1.0045	&	74.5	&	<0.008	&	-	\\
		J1509+5531	&	0.7396	&	19.6	&	<0.002	&	-	&	J2215+1538	&	0.3741	&	29.2	&	0.023	&	$120 \pm 18$	\\
		J1532+2745	&	1.1248	&	14.6	&	<0.003	&	-	&	J2227+3038	&	0.8424	&	19.9	&	<0.005	&	-	\\
		J1538+2345	&	3.4493	&	14.9	&	<0.015	&	-	&	J2234+2114	&	1.3587	&	35	&	<0.005	&	-	\\
		J1543+0929	&	0.7484	&	34.9	&	<0.002	&	-	&	J2253+1516	&	0.7922	&	29.2	&	<0.013	&	-	\\
		J1549+2113	&	1.2624	&	24	&	<0.013	&	-	&	J2305+3100	&	1.5758	&	49.5	&	0.01	&	$189 \pm 5$	\\
		J1614+0737	&	1.2068	&	21.3	&	<0.004	&	-	&	J2306+31	&	0.3416	&	46.1	&	<0.007	&	-	\\
		J1627+1419	&	0.4908	&	32.1	&	<0.004	&	-	&	J2317+2149	&	1.4446	&	20.8	&	0.004	&	$279 \pm 5$	\\
		\hline
	\end{tabular}
	\label{tab:tab2}
\end{table*}

Out of a total of 96 pulsars in Table 1, $IP$ were detected in 12 pulsars. Fig.~\ref{fig:fig1A} shows the summarized average profiles of these pulsars. The $IP$ phase for 7 out of 12 pulsars, taking into account the above errors, differs from $180^{\circ}$ by no more than $15^{\circ}$. This means that pulsars J0613+3731, J0826+2637, J0953+0755, J1946+1805, J1952+1510, J2208+4056, J2305+3100 can be orthogonal rotators. Polarizing observations are necessary to determine whether they are actually orthogonal or coaxial rotators. Unfortunately, only one linear polarization can be observed on the LPA radio telescope, and we cannot make such observations. Of the 12 pulsars with $IP$ found, three pulsars (J0826+2637, J0953+0755, J1946+1805) had $IP$ previously detected by other observers. According to \citeauthor{Maciesiak2011} (\citeyear{Maciesiak2011}), J0826+2637 is an orthogonal rotator, while J0953+0755 and J1946+1805 are coaxial rotators. 

The $IP$ phase for 5 of the 12 pulsars is far from $180^{\circ}$. That is, pulsars J1823+0550, J2018+2839, J2113+2754, J2215+1538, J2317+2149 should be coaxial rotators. The proportion of orthogonal and coaxial rotators in the part of pulsars with $IP$ cannot be determined, but given that J0953+0755 and J1946+1805 are coaxial rotators, the proportion of coaxial rotators in the total sample of 12 pulsars with $IP$ found exceeds 58\%.

In the paragraph "Observations and processing", it was said that we should detect at least 90\% of all pulsars with $IP$ having a flux density at a maximum of $IP$ of at least 10\% of the peak density of $MP$. However, we have not been able to detect any pulsars with such a high value of $IP/MP$. The found values of $IP/MP$ lie between 0.4\% (J2018+2839; J2317+2149) and 2.3\% (J2215+1538), the median value is 1.0-1.1\%. At the same time, for the remaining part of the sample of 84 pulsars without $IP$ found, the median upper estimate of the ratio of $IP$ to $MP$ is 0.7\%, the maximum estimate is 8.7\% (J2017+2043), the minimum estimate is 0.01\% (J1921+2153).

\section{Discussion of the results and conclusion}

The selection effect that dramatically reduces the sensitivity of the LPA and affects the total detectable number of pulsars with $IP$ is interstellar scattering. In addition to this effect, the sensitivity is also affected by dispersion smoothing in frequency channels, the sampling of the point, the height of the source at the time of passing the central meridian, industrial and other interference.

In the meter wavelength range, the effect of interstellar scattering, in addition to worsening the sensitivity of observations, makes it impossible to observe pulsars with small periods, since the characteristic scattering time may be longer than the period. According to the work \citeauthor{Maciesiak2011} (\citeyear{Maciesiak2011}); \citeauthor{Tyulbashev2024} (\citeyear{Tyulbashev2024}), where the practical sensitivity of LPA3 was tested when searching for pulsars, it turned out that the best sensitivity is achieved for pulsars with $+21^{\circ} < \delta < +42^{\circ}$, $0.25 <P < 3$s, $DM < 100$~pc/cm$^3$.

At declinations $\delta < +21^{\circ}$, there is more industrial interference in the recordings, since low declinations correspond to sources close to the horizon, and interference comes to the radio telescope from large industrial centers located in the cities of Moscow and Tula. The effective area of the radio telescope decreases as the cosine of the source height. Sensitivity decreases at periods of $P < 0.25$s, because at such periods the pulse duration is comparable to the sampling of the point, and already at $DM > 25$~ pc/cm$^3$ in the 78~kHz frequency channel, the pulse broadening due to dispersion smoothing is comparable to the sampling of the point.

In addition to these factors, the scattering effect, proportional to the fourth power of the frequency of observations, smoothes the pulses and thereby reduces the sensitivity of observations. Observations show that the scattering effect for second pulsars according to observations at 102 MHz \citeauthor{Kuzmin2007} (\citeyear{Kuzmin2007}); \citeauthor{Pynzar2008} (\citeyear{Pynzar2008}) becomes significant at $DM > 70-80$~pc/cm$^3$. At $DM > 250$~pc/cm$^3$ there is not a single pulsar visible on the LPA (https://bsa-analytics.prao.ru/pulsars/). Therefore, for various reasons, we do not see many pulsars, which, all other things being equal, are visible at high frequencies.

However, we have no reason to assume that the proportion (percentage) of pulsars with $IP$ may be different for pulsars with different $DM$. Different heights above the horizon when the pulsar passes through the meridian and/or any kind of interference can lead to a decrease in the total number of pulsars observed, but the effect of these factors is the same on pulsars without or with interpulses.

In our opinion, the only selection effect acting differently on pulsars with and without interpulses is associated with observations of pulsars with small periods. As can be seen from the observations of interpulses, $IP$ can be located at different phase distances with respect to $MP$. Considering that the scattering at any measure of dispersion can differ by 1-1.5 orders of magnitude from the average (expected) value of \citeauthor{Kuzmin2007} (\citeyear{Kuzmin2007}), it can "absorb" interpulse and this absorption will be effective, especially for coaxial rotators when the interpulse is close to the main pulse. $MP$ can be 100 or more times stronger than $IP$, therefore, the weaker the interpulse, the more likely it is to disappear due to scattering. These factors may lead to a decrease in the proportion of observed pulsars with $IP$ compared to conditions when they are not active. This selection effect will be especially strong for pulsars with periods of $P < 0.25$s and $DM > 100$~pc/cm$^3$. In the paragraph "Observations and processing" it is said that the search for pulsars with $IP$ was carried out for pulsars with $P > 0.25$s. Pulsars with $DM > 100$~pc/cm$^3$ in the sample are only 2 out of 96, and therefore the effect of the scattering effect, which can lead to the loss of pulsars with $IP$, should not be significant.

Before comparing our results with the results of other works, both observational (\citeauthor{Weltevrede2008}, \citeyear{Weltevrede2008}; \citeauthor{Wang2023}, \citeyear{Wang2023}; \citeauthor{Posselt2023}, \citeyear{Posselt2023}) and theoretical (\citeauthor{Arzamasskiy2017}, \citeyear{Arzamasskiy2017}; \citeauthor{Novoselov2020}, \citeyear{Novoselov2020}), we note two more points that may affect the discussed statistics of interpulse pulsars.

First, almost all the interpulse pulsars we observed in Table 1 have an intensity ratio of $IP/MP < 0.02$, whereas at least half of all interpulse pulsars from other catalogs have an intensity ratio of $IP/MP > 0.1$. This point has no explanation yet. The search for pulsars with $IP$ was conducted in the site $-9^{\circ} < \delta < +55^{\circ}$, it covers almost half of the entire sky. In the work  \citeauthor{Arzamasskiy2017} (\citeyear{Arzamasskiy2017}), 12 pulsars with a ratio of $IP/MP > 0.1$ can be seen falling into the studied site, and at the same time missing from Table 1. Of these, 11 pulsars (J0627+0706, J1842+0358, J1843-0702, J1849+0409, J1851+0418, J1852-0118, J1903+0925, J1913+0832, J1915+1410, J2032+4127, J2047+5029) are in the PUMPS survey (https://bsa-analytics.prao.ru/pulsars/known/) not detected. 10 of these 11 pulsars have $DM > 100$~pc/cm$^3$. As we have already noted, the sensitivity of the survey on large measures of variance is low. Half of these 10 pulsars, in addition to the high $DM$, also have a period of $P < 0.25$s. One pulsar (J1849+0409) has $P = 0.7611$s, $DM= 63.9$~pc/cm$^3$, and should be detected in PUMPS (\citeauthor{Tyulbashev2022}, \citeyear{Tyulbashev2022}). However, with the achieved survey sensitivity of 0.1-0.2 mJy for the integral flux density, the harmonics of the pulsar J1849+0409 were not found in the averaged power spectra. Finally, one pulsar J0535+2200 (pulsar in the Crab) was detected by us both in individual pulses and in averaged power spectra. However, its period is only 33 ms, that is, less than 3 points of 12.5 ms, so it is impossible to detect its $IP$ in our observations. Nevertheless, the question is why almost all interpulse pulsars with $IP/MP > 0.1$ and entering the studied site have high $DM$, and the question is why pulsars with $DM < 100$~pc/cm$^3$ with $IP/MP$ have not been found $> 0.1$, have no answers.

Note that it is not the predominance of interpulse pulsars with low $IP/MP$ ratios that is surprising here, but the relatively small number of pulsars with $IP/MP \sim 1$. Indeed, the low ratio of $IP/MP$ can easily be explained if we assume a significant difference in the angle between the magnetic moment and the axis of rotation from $90^\circ$. In this case, the beam will cross the radiation pattern from one of the magnetic poles away from the magnetic axis. The ratio $IP/MP = 1$ should be performed only for tilt angles close to $90^\circ$.

Secondly, all statistical distributions discussed in \citeauthor{Arzamasskiy2017} (\citeyear{Arzamasskiy2017}); \citeauthor{Novoselov2020} (\citeyear{Novoselov2020}) relate to pulsars with an intensity ratio of $IP/MP > 0.01$. Therefore, when comparing the number of interpulse pulsars below, we will limit ourselves to only seven pulsars from Table 1 having the same parameters. Of these, four pulsars (J0613+3731, J0826+2637, J2208+4056, J2305+3100) can be attributed to orthogonal inte pulse pulsars (the angular distance between $MP$ and $IP$ is close to 180 degrees), and three pulsars (J0953+0755, J1946+1805, J2215+1538) it can be attributed to coaxial rotators.

\begin{table*}
\centering
\caption{Observed number of orthogonal interpulse pulsars for three period intervals $P$}
\label{tab:tab1}
\begin{tabular}{c|c|c|c|c|c|c}
\hline
P(s) & $N_{all}$ & $N_{IP}$ & N(\%) & \citeauthor{Novoselov2020} (\citeyear{Novoselov2020}) (\%) & FAST (\%) & MeerKAT (\%) \\
\hline
0.03-0.5 & 26 & 1-2 & 1.0-2.1 & 2.2-3.6 & 2.0 & 2.4\\
0.5-1.0 & 28 &  0-1 & 0-1.0 & 0.7-1.1 & 0.4 & 0.5\\
>1 & 42 & 1 & 1.0 & $\le$0.2 & 0.3 & 0.2\\
\hline
\end{tabular}
\label{tab:tab2}
\end{table*}

Table~\ref{tab:tab2} shows the obtained values of the number of orthogonal interimpulse pulsars for three intervals of periods $P$, as well as the values of the total number of pulsars and pulsars with interpulses (columns 2 and 3). Column 4 shows the relative number of interpulse pulsars taken from (\citeauthor{Novoselov2020}, \citeyear{Novoselov2020}), and columns 5 and 6 from surveys of FAST (\citeauthor{Wang2023}, \citeyear{Wang2023}) and MeerKAT (\citeauthor{Posselt2023}, \citeyear{Posselt2023}). The uncertainty is due to the fact that the pulsar PSR J0826+2637 has a period of $P = 0.53$s, i.e. it is located on the very border of the two intervals. In general, as we see, there is a good agreement between our results and those of other authors.

Recall that in \citeauthor{Novoselov2020} (\citeyear{Novoselov2020}) calculations were performed estimating the expected relative number of observed orthogonal interimpulse pulsars for two models of neutron star deceleration. In the first, the so-called MHD model (see, for example, (\citeauthor{Spitkovsky2006}, \citeyear{Spitkovsky2006}; \citeauthor{Philippov2014}, \citeyear{Philippov2014}), in which the angle between the magnetic axis and the axis of rotation tends to $0^{\circ}$, the number of orthogonal interpulse pulsars in the range of $0.033<P<0.5$s at the level of 0.2 is predicted-1.5\%. In the second, BGI-model (\citeauthor{Gurevich1993}, \citeyear{Gurevich1993}), in which the angle between the magnetic axis and the axis of rotation tends to 90 degrees, the relative amount of IP is predicted at the level of 2.5-5.5\%.

Thus, we can draw a preliminary conclusion that our results (as well as the latest results of the FAST and MeerKAT observatories) speak more in favor of the BGI model than in favor of the generally recognized MHD model. As for coaxial interpulse pulsars, as shown in \citeauthor{Arzamasskiy2017} (\citeyear{Arzamasskiy2017}), both the MHD model and the BGI model predict approximately the same relative number of them at the level of 1-2\%, which is generally consistent with our results (3 pulsars with $IP/MP > 0.01$ out of 96).

In conclusion, we note the main results obtained in this work:

- for 96 pulsars observed in the PUMPS monitoring survey, by adding up the average profiles, it was possible to obtain minimum $S/N$ in the main pulse of more than 40. To obtain the average profiles, observational sessions were used in which the $S/N$ of the main pulse was greater than 6. These sessions were selected from daily observations lasting over time intervals of hundreds to thousands of days;

- 12 pulsars with $IP$ have been found. 7 pulsars have $IP$ located at phase distances close to $180^{\circ}$ relative to $MP$. For 5 pulsars, the distances are far from $180^{\circ}$ and they appear to be coaxial rotators;

- for pulsars with a ratio of $IP/MP > 0.01$, the results do not contradict the model from \citeauthor{Novoselov2020} (\citeyear{Novoselov2020}), which assumes the evolution of pulsars into orthogonal rotators.

\section*{ACKNOWLEDGMENTS}

The study was carried out at the expense of a grant Russian Science Foundation 22-12-00236, https://rscf.ru/project/22-12-00236/. The authors thank L.B. Potapova for help in preparing the paper.



\end{document}